\documentclass{jetpl}
\twocolumn

\lat
\usepackage{graphicx}

\title {Applicability of Anderson and Hubbard model for Ce metal 
and cerium  heavy fermion compounds}
\rtitle {Applicability of Anderson and Hubbard model \ldots}
\sodtitle {Applicability of Anderson and Hubbard model for Ce metal 
and cerium  heavy fermion compounds}

\author {S.\,V.\,Streltsov, A.\,O.\, Shorikov, V.\,I.\,Anisimov}
\rauthor {S.\,V.\,Streltsov, A.\,O.\, Shorikov, V.\,I.\,Anisimov}
\sodauthor {Streltsov, Shorikov, Anisimov}

\address {Institute of Metal Physics, Russian Academy of Sciences, 
620041 Yekaterinburg GSP-170, Russia \\~\\
Ural Federal University, Mira St. 19, 620002 
Ekaterinburg, Russia}

\dates {\today}{*}

\abstract{ 
The importance of taking into account inter-site $f-f$ hybridization in electron structure 
calculations for Ce metal and cerium heavy fermion compounds was studied. 
We demonstrate that for heavy-fermion systems such as cerium 
compound CeCu$_2$Si$_2$ $f-f$ hybridization can be neglected and Anderson model application 
is well justified. On another hand for cerium metal $f-f$ hybridization is 
strong enough to provide the contribution to hybridization 
function comparable to hybridization between  $4f$ and itinerant electrons. 
We argue that in the case of Ce only the most general Hamiltonian combining 
Hubbard and Anderson models should be used.}


\begin{document}

\maketitle

The mysterious properties of metallic Ce, which has paramagnetic
phase with the local magnetic moments at ambient pressure and
room temperature (Ce$-\gamma$ phase), and show the absence
of local moments and Pauli paramagnetism below $\sim$100 K (Ce$-\alpha$ phase)
rivet attention of the researchers~\cite{Koskenmaki-78}.
For decades the electronic and magnetic properties of metallic 
Ce and heavy fermion cerium compounds were considered in the frameworks of the single impurity problem mainly using
Anderson impurity model~\cite{Anderson-61}:
\begin{eqnarray}
 \nonumber
 \hat H_{SIAM}=\sum_{\bf k\sigma}\varepsilon_{\bf k}\hat{c}^{\dagger}_{\bf k\sigma}\hat{c}_{\bf k\sigma}+\varepsilon_{f}\sum_{\sigma}\hat{f}^{\dagger}_{\sigma}\hat{f}_{\sigma}
 +U\hat{n}_{f\uparrow}\hat{n}_{f\downarrow}+
 \\
 \sum_{\bf k\sigma}
 \left(
  V_{\bf k}\hat{c}^{\dagger}_{\bf k\sigma}\hat{f}_{\sigma}+V^{*}_{\bf k}\hat{f}^{\dagger}_{\sigma}\hat{c}_{\bf k\sigma}
 \right) \ .
 \label{HSIAM}
\end{eqnarray}
where localized $f$-electrons with on-cite Coulomb interaction term $U\hat{n}_{f\uparrow}\hat{n}_{f\downarrow}$ hybridize with itinerant $c$-electrons described by dispersion $\varepsilon_{\bf k}$ with a hybridization strength parameter $V_{\bf k}$. 

One can introduce noninteracting Green function ${\cal G}_{0}$ (defined as Green function  with Coulomb 
interaction switched off):
\begin{eqnarray} 
\label{Green-12z}
{\cal G}_{0}(i\omega_n)=(i\omega_n+\mu-\epsilon_d-\Delta(i\omega_n))^{-1} \ ,
\end{eqnarray}
where $\omega_n=(2n+1)\pi T$, $n=0,\pm 1,\pm 2,\ldots$ are Matsubara frequencies and
hybridization function $\Delta(i\omega_n)$ is defined as:
\begin{eqnarray} 
\label{Green-12l}
\Delta(i\omega_n)=\sum_{{\bf k}}\frac{|V_{\bf k}|^2}{i\omega_n-\epsilon_{{\bf k}}+\mu}.
\end{eqnarray}

Then the problem that should be solved is to describe $f$-electrons with on-cite 
Coulomb interaction in an effective media defined by noninteracting Green 
function ${\cal G}_{0}$ (\ref{Green-12z}) where interaction with effective media is 
determined by hybridization function $\Delta(i\omega_n)$ (\ref{Green-12l}).

The calculations performed using this model allowed to obtain consistent 
description of the evolution of magnetic and electronic properties as due to
appearance of the Kondo scattering in $\alpha-$phase of Ce.
The impurity models were applied for the study of the magnetic 
susceptibility~\cite{Rajan-83},  specific heat~\cite{Rajan-83} 
and different types of spectra (photoemission~\cite{Patthey-90}, Bremsstrahlung 
isochromatic~\cite{Liu-92}, electron-energy-loss~\cite{Wuilloud-83}). 
Fitting of the theoretical result obtained within impurity models to 
different experimental data (protoemission spectra, 
susceptibility etc.) allows to extract the most important parameters 
in Kondo physics - Kondo temperature $T_K$~\cite{Liu-92}.

While Anderson impurity model (\ref{HSIAM}) has allowed to capture main energy scale in 
heavy-fermion physics - Kondo temperature $T_K$, it cannot describe coherence effects  
when at low temperatures rich phase diagram appears with long-range magnetic ordering 
and superconductivity.
Basic model used to describe such effects for $f$-systems is periodic 
Anderson model ($PAM$) with Hamiltonian:
\begin{eqnarray}
  \nonumber \hat H=\varepsilon_{c}\sum_{i\sigma}\hat{c}^{\dagger}_{i\sigma}\hat{c}_{i\sigma}+
  \sum_{ij\sigma}t_{ij}\hat{c}^{\dagger}_{i\sigma}\hat{c}_{j\sigma}+ 
  \varepsilon_{f}\sum_{i\sigma}\hat{n}^{f}_{i\sigma}+ \\
  U\sum_{i}\hat{n}^{f}_{i\uparrow}\hat{n}^{f}_{i\downarrow}+ \sum_{ij\sigma}\left(V_{ij}\hat{c}^{\dagger}_{i\sigma}\hat{f}_{j\sigma}+V^{*}_{ij}\hat{f}^{\dagger}_{j\sigma}\hat{c}_{i\sigma}\right).
 \label{HPAM}
\end{eqnarray}
It deals with localized  $f$-electrons on all sites embedded in 
itinerant $c$-electrons bath with a term responsible for hybridization between 
localized and itinerant electrons.

In both impurity (\ref{HSIAM}) and periodic (\ref{HPAM}) Anderson models 
hybridization between $f$-electrons on different lattice sites is assumed 
to be absent in contrast to Hubbard model where competition between 
inter-site $f-f$ hybridization and Coulomb on-site interaction is 
explicitly defined:
\begin{eqnarray}
\label{Hubbard}
  \nonumber 
  \hat H=\sum_{ij\sigma}t^f_{ij}\hat{f}^{\dagger}_{i\sigma}
  \hat{f}_{j\sigma}+  \varepsilon_{f}\sum_{i\sigma}\hat{n}^{f}_{i\sigma}+
  U\sum_{i}\hat{n}^{f}_{i\uparrow}\hat{n}^{f}_{i\downarrow}.
\end{eqnarray}

\begin{figure}[t]
 \centering
 \includegraphics[clip=false,angle=270,width=0.58\textwidth]{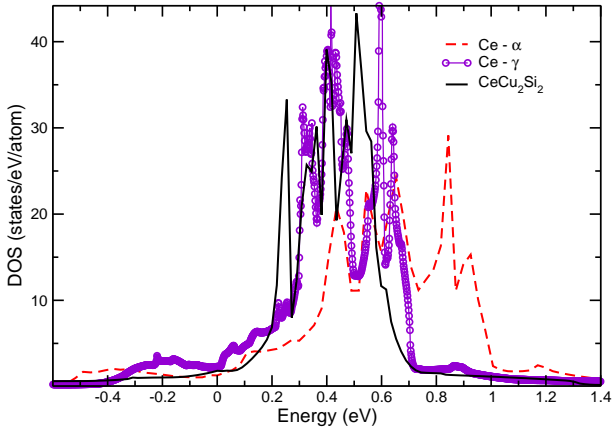}
\caption{\label{DOS}Fig.~\ref{DOS} (Color online). LDA Ce$-4f$ partial DOS for 
Ce-$\alpha$ (dashed, red), Ce-$\gamma$ (solid with circles, violet) and 
CeCu$_2$Si$_2$ (solid, black). The Fermi level is in zero.}
\end{figure}

If one cannot neglect inter-site $f-f$ hybridization then the most general Hamiltonian combining 
Hubbard and Anderson models should be defined and studied:
\begin{eqnarray}
  \nonumber \hat H=\varepsilon_{c}\sum_{i\sigma}\hat{c}^{\dagger}_{i\sigma}\hat{c}_{i\sigma}+
  \sum_{ij\sigma}t_{ij}\hat{c}^{\dagger}_{i\sigma}\hat{c}_{j\sigma}+ \\
  \nonumber
  \sum_{ij\sigma}t^f_{ij}\hat{f}^{\dagger}_{i\sigma}\hat{f}_{j\sigma} + 
  \varepsilon_{f}\sum_{i\sigma}\hat{n}^{f}_{i\sigma}+U\sum_{i}\hat{n}^{f}_{i\uparrow}
  \hat{n}^{f}_{i\downarrow}+ \\ 
  \sum_{ij\sigma}\left(V_{ij}\hat{c}^{\dagger}_{i\sigma}
  \hat{f}_{j\sigma}+V^{*}_{ij}\hat{f}^{\dagger}_{j\sigma}\hat{c}_{i\sigma}\right).
 \label{HPAM2}
\end{eqnarray}

In the present paper we investigate the problem of applicability of Anderson model to study cerium and cerium compounds and estimate the strength of inter-site $f-f$ hybridization. We demonstrate that while for heavy-fermion systems such as cerium compound CeCu$_2$Si$_2$ $f-f$ hybridization can be neglected and Anderson model (\ref{HPAM}) application is well justified, for cerium metal 
inter-site $f-f$ hybridization is strong enough giving contribution 
to hybridization function (\ref{Green-12l}), which is comparable to hybridization of  
$f$-electrons with itinerant electrons. In the last case only the most general 
Hamiltonian (\ref{HPAM2}) should be used.

With the use of the Linear muffin-tin orbitals 
(LMTO) method~\cite{Andersen-84} and the Local density 
approximation (LDA) we show that Ce-$4f$ states in metallic Ce 
should not be described simply as impurity levels. These states 
do form bands and $f-f$ hopping matrix elements between different
Ce sites are sizable. In contrast the $f$-states in Ce compounds
are more localized and do not show 
significant band dispersion.

We start from the Ce-$4f$ partial Density of states (DOS) for 
Ce-$\alpha$, Ce-$\gamma$ and CeCu$_2$Si$_2$ presented
in Fig.~\ref{DOS}. One may see that the widths of the DOS are 
comparable for all three systems and hence it may be expected 
that the band characteristics of f-states in these compounds
are similar. Since the similarity in the position, width and shape of partial DOS 
is most pronounced for Ce-$\gamma$ and CeCu$_2$Si$_2$ we will
use these two systems to compare band effects.

\begin{figure}
 \centering
 \includegraphics[clip=false,angle=270,width=0.5\textwidth]{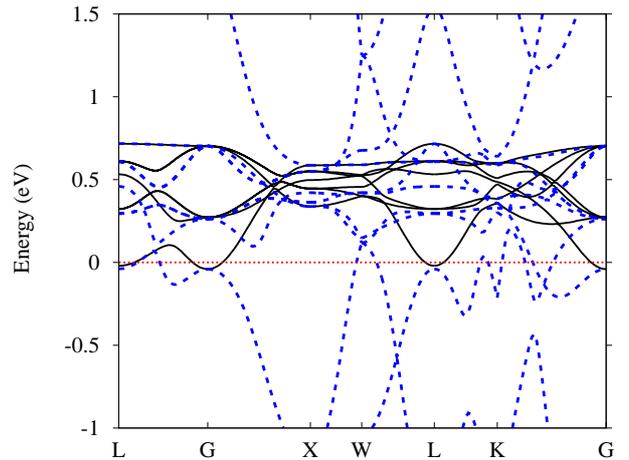} 
\caption{\label{Bands-Ce} Fig.~\ref{Bands-Ce} (Color online). Full orbital LDA band structure 
for Ce-$\gamma$ is shown by blue dashed curve. 
The band structure obtained by removing Ce$-s,p,d$ states from 
the self-consistent LDA hamiltonian is shown in black. They are more localized,
but still cannot be considered as atomic levels. The Fermi level
corresponds to zero energy.
}
\end{figure}

The real band structures obtained in the self-consistent LDA
calculation for Ce-$\gamma$ is shown by dashed 
curves in Fig.~\ref{Bands-Ce}. Seven Ce-$4f$ are spread over wide 
energy [-0.4 eV, 1 eV] (compare with Fig.~\ref{DOS}). 

In order to check whether Ce-$4f$ states can be treated as independent 
impurity states we remove (set zero) all the matrix elements 
from the self-consistent LDA hamiltonian except Ce-$4f$. 
The self-consistent potential for the real material is still used, so 
that the resulting band structure is {\it not} the same as for hypothetical
``Ce-f-only ions'' in Ce-$\gamma$ type lattice. The band structure obtained 
within this method can be thought as the 
actual dispersion of Ce$-f$ states in real Ce$-\gamma$, where 
hybridization with Ce$-s,p,d$ states was switched off.
The same procedure was previously applied for the analysis of the 
chemical bonding in Ag$_2$NiO$_2$~\cite{Johannes-07}. 

The comparison of full-orbital LDA bands structure and 
one obtained removing Ce-$s,p,d$ states from the basis set is
shown in Fig.~\ref{Bands-Ce}. One may see that the 
band dispersion of Ce-$4f$ states is quite similar, and
that these states still form the real bands, rather than 
atomic levels. The band-width W$_{f-only}$ $\sim$ 0.75~eV
in Ce$-\gamma$.

\begin{figure}
 \centering
 \includegraphics[clip=false,angle=270,width=0.5\textwidth]{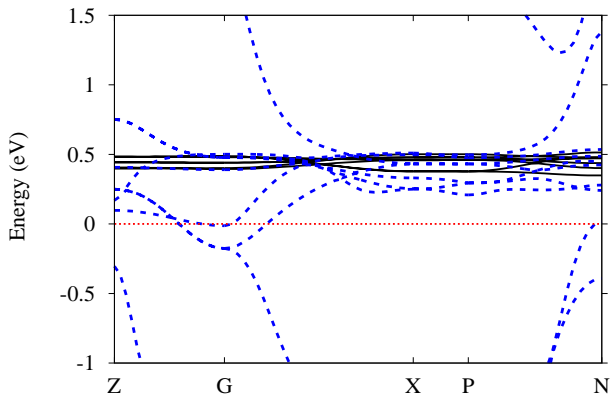}
\caption{\label{Bands-CeCuSi} Fig.~\ref{Bands-CeCuSi} (Color online).
Full orbital LDA band structure for CeCu$_2$Si$_2$ is shown by blue dashed curve. 
The band structure obtained by removing Ce$-s,p,d$ states from 
the self-consistent LDA hamiltonian is shown in black. The Fermi level
corresponds to zero energy.}
\end{figure}

In order to show that this situation is specific to Ce we
performed the same calculations for 
CeCu$_2$Si$_2$. The results are presented in Fig.~\ref{Bands-CeCuSi}.
In contrast to the case of metallic Ce the absence of the hybridization
between Ce$-4f$ and Ce-$s,p,d$ states leads to the loss of
band dispersion. The reason for such a different behavior of
metallic Ce and CeCu$_2$Si$_2$ is rather obvious: in the last
case Ce ions are separated by Cu and Si, direct $f-f$ hopping
and corresponding effective bandwidth is small (W$_{f-only}$ $\sim$ 0.1~eV) 
and the bands are dispersionless like atomic levels. However,
the presence of sizable band dispersion for metallic Ce was not
taken into account in previous model 
calculations.

The value of the Ce $f-f$ hopping parameters estimated from
the band-width and tight-binding parametrization or more
sophisticated Wannier projection procedure~\cite{Streltsov-05}
results in $t_{ff} \sim$30 meV. The presence of small, but finite $f-f$ hopping may lead
to a number of consequences. The most obvious is a direct antiferromagnetic
exchange interaction between Ce ions proportional to $2t^2_{ff}/U$. Together 
with indirect Ruderman-Kittel-Kasuya-Yosida (RKKY) exchange this interaction 
will act against formation of a coherent state. 

The most direct 
investigation of the effects connected with the presence of finite $f-f$ hoppings in pure Ce 
can be performed by a numerical solution of~(\ref{HPAM2}), using for instance
Dynamical mean-field theory (DMFT) or its cluster extension~\cite{Georges-96}.
However, already on the LDA level one may show that these effects 
should be important. In order to demonstrate it the hybridization 
function on the real energy axis was constructed with and without
$f-f$ hopping. On the first step of this procedure one obtains
Hamiltonian in the basis of Wannier functions in real space 
as described in ref.~\cite{Streltsov-05}. Then one makes zero corresponding
off-diagonal matrix elements, perform back Fourier transform to reciprocal
space, calculate density of states using tetrahedron method and
construct hybridization function $\Delta(\varepsilon)$ using formalism developed
in ref.~\cite{Gunnarsson-89}:
\begin{eqnarray}
\label{Hyb}
\Delta(\varepsilon) = - Im \sum_{\nu} \Big( \int \frac{\rho_{\nu}(\varepsilon')}
 {\varepsilon - \varepsilon' - i \theta} d\varepsilon' \Big)^{-1},   
\end{eqnarray}
where $\rho_{\nu}(\varepsilon)$ is a partial DOS, and $\nu$ - orbital index. 
To avoid numerical errors partial DOS were normalized on unity before 
apply~(\ref{Hyb}).

\begin{figure}
 \centering
 \includegraphics[clip=false,angle=270,width=0.5\textwidth]{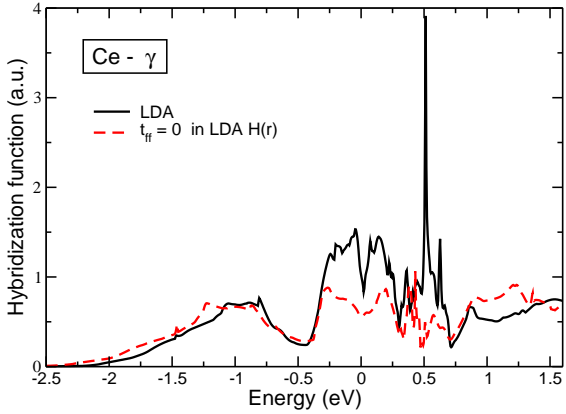}
\caption{\label{Hyb-Ce} Fig.~\ref{Hyb-Ce} (Color online).
Total hybridization function $\Delta(\varepsilon)$ for Ce$-\gamma$ as defined
in~(\ref{Hyb}) calculated in conventional LDA and in LDA, where in self-consistent Hamiltonian
off-diagonal inter-site $f-f$  matrix elements were put to zero. 
The Fermi level corresponds to zero energy.}
\end{figure}

The plot of the hybridization function obtained in this way in comparison
with $\Delta (\varepsilon)$ from conventional LDA calculation is presented in 
Fig.~\ref{Hyb-Ce}. One may see that the most significant changes in frequency
are observed near the Fermi level. The full description of the electronic properties
of the system with given hybridization can be obtained only by numerical
solution of many-body problem. However, already on the LDA level we obtain that
the ratio $\bar \Delta_{LDA}(\varepsilon) / \bar \Delta_{t_{ff}=0}(\varepsilon)$ is of order 2
for Ce and 1.2 for CeCu$_2$Si$_2$, where $\bar \Delta (\varepsilon)$ is 
averaged over the region of 1 eV around the Fermi level total hybridization function.
This demonstrates an importance of the account of direct $f-f$ hopping matrix 
elements in a real many-body calculation

To sum up, in the present paper we've shown that there is
sizable $f-f$ hopping matrix element in the metallic Ce.
This implies that the full description of the electronic 
properties of Ce should be obtained not within the frameworks of the single impurity, 
but rather in lattice models, where hopping parameters between different
f-sites are implicitly taken into account. Thus, multi-band Hubbard model
is one of the models suitable for such an investigation.


This work was supported by grants RFBR 10-02-00046 and 10-02-96011,
the program of President of Russian Federation MK-309.2009.2,
the Russian Federal Agency of Science and Innovation N 02.740.11.0217,
the scientific program ``Development of scientific potential of universities'' 
N 2.1.1/779, grant UB and SB of 
RAS $\mathcal N^{\circ \!\!\!\!\_~}$22.


\begin{thebibliography}{99}
\bibitem{Koskenmaki-78} See review D.\,C. Koskenmaki and K.\,A. Gschneider, Jr.
in Handbook on the Physics and Chemistry of the Rare Earths, edited by 
K.\,A. Gschneider and L. Eyring (North-Holland, Amsterdam, 1978), Vol. 1, Chap. 4.
\bibitem{Anderson-61} P.\,W. Anderson, Phys.~Rev. {\bf 124}, 41, (1961).
\bibitem{Rajan-83} V.T. Rajan, Phys.~Rev.~Lett {\bf 51}, 308, (1983).
\bibitem{Patthey-90} F. Patthey, J.-M. Imer,  W.-D. Schneider, H. Beck,
Y. Baer and B. Delley, Phys.~Rev.~B {\bf 42}, 8864, (1990).
\bibitem{Liu-92} L.\,Z. Liu, J.\,W. Allen, O. Gunnarsson, N.\,E. Christensen, and 
O.\,K. Andersen, Phys.~Rev.~B {\bf 45}, 8934, (1992).
\bibitem{Wuilloud-83} E. Wuilloud, H.\,R. Moser, W.-D. Schneider, and Y. Baer,
Phys.~Rev.~B {\bf 28}, 7354, (1983).
\bibitem{Andersen-84} O.\,K. Andersen and O. Jepsen,
Phys.~Rev.~Lett. {\bf 53}, 2571 (1984). 
\bibitem{Johannes-07} M.\,D. Johannes, S.\,V. Streltsov, I.\,I. Mazin and D.\,I. Khomskii,
 Phys.~Rev.~B {\bf 75}, 180404, (2007).
\bibitem{Streltsov-05} S.\,V. Streltsov, A.\,S. Mylnikova,
A.\,O. Shorikov, Z.\,V. Pchelkina, D.\,I. Khomskii, and V.\,I. Anisimov,
Phys.~Rev.~B {\bf 71}, 245114 (2005).
\bibitem{Georges-96} A. Georges, G. Kotliar, W. Krauth and M.\, J. Rozenberg,
Rev.~Mod.~Phys. {\bf 68}, 13 (1996).
\bibitem{Gunnarsson-89} O. Gunnarsson, O.\,K. Andersen, O. Jepsen and J. Zaanen, 
Phys.~Rev.~B {\bf 39}, 1708 (1989).

\end{thebibliography}
\end{document}